# A Self-Decoupled 32 Channel Receive Array for Human Brain Magnetic Resonance Imaging at 10.5T

Nader Tavaf[1], Russell L. Lagore[1], Steve Jungst[1], Shajan Gunamony[2], Jerahmie Radder[1], Andrea Grant[1], Steen Moeller[1], Edward Auerbach[1], Kamil Ugurbil[1], Gregor Adriany[1], Pierre-Francois Van de Moortele[1]

[1.] Center for Magnetic Resonance Research (CMRR), University of Minnesota, Minneapolis, MN
[2.] Centre for Cognitive Neuroimaging, University of Glasgow, Glasgow, Scotland

* Correspondence: Nader Tavaf, Email: tavaf001@umn.edu

**Purpose**: Receive array layout, noise mitigation and $B_0$ field strength are crucial contributors to signal-to-noise ratio (SNR) and parallel imaging performance. Here, we investigate SNR and parallel imaging gains at 10.5 Tesla (T) compared to 7T using 32-channel receive arrays at both fields.

**Methods**: A self-decoupled 32-channel receive array for human brain imaging at 10.5T (10.5T-32Rx), consisting of 31 loops and one cloverleaf element, was co-designed and built in tandem with a 16-channel dual-row loop transmitter. Novel receive array design and self-decoupling techniques were implemented. Parallel imaging performance, in terms of SNR and noise amplification (g-factor), of the 10.5T-32Rx was compared to the performance of an industry-standard 32-channel receiver at 7T (7T-32Rx) via experimental phantom measurements.

**Results**: Compared to the 7T-32Rx, the 10.5T-32Rx provided 1.46 times the central SNR and 2.08 times the peripheral SNR. Minimum inverse g-factor value of the 10.5T-32Rx (min(1/g) = 0.56) was 51% higher than that of the 7T-32Rx (min(1/g) = 0.37) with R=4x4 2D acceleration, resulting in significantly enhanced parallel imaging performance at 10.5T compared to 7T. The g-factor values of 10.5T-32Rx were on par with those of a 64-channel receiver at 7T, e.g. 1.8 versus 1.9, respectively, with R=4x4 axial acceleration.

**Conclusion**: Experimental measurements demonstrated effective self-decoupling of the receive array as well as substantial gains in SNR and parallel imaging performance at 10.5T compared to 7T.

**Keywords:**
RF coils, self-decoupling, receive array, parallel imaging, ultra-high field MRI, noise correlation.

## 1. INTRODUCTION

Ultrahigh field (UHF) magnetic resonance imaging (MRI) advantages, including higher image resolution, reduced acquisition time, and better signal-to-noise ratio (SNR), have opened new opportunities for advancement of various clinical and basic research projects (e.g. [1–4]). Applications of primary interest that rely on exploiting these UHF advantages include functional MRI (fMRI) studies (e.g. [2,4–7]), brain connectivity mapping using diffusion-weighted MR imaging (dMRI) (e.g. [8–11]), and anatomical imaging (e.g. [12–15]).

The push for UHF MRI is predicated primarily on the premise of significant ultimate intrinsic SNR[16–19] gains and less SNR penalty with parallel imaging acceleration[13,20–23] at higher $B_0$ field strengths. However, UHF MRI presents significant new technical challenges[1,2,24–26]; a critically important technology that enables acquisition strategies that addresses some of these challenges, such as reducing echo-train lengths in fast acquisition schemes like echo planar imaging (EPI), is parallel imaging[22] using high density receive arrays[23,27–29].

Acceleration via parallel imaging comes with a penalty in SNR quantified in terms of noise amplification or g-factor[30]. Increasing the $B_0$ field strength has been shown to mitigate the SNR penalty attributed to acceleration[22,31,32], hence affording better parallel imaging performance. However, parallel imaging performance depends significantly on receiver array noise correlation[33,34] as well. Increasing receive array density, on the other hand, exacerbates interelement coupling (noise correlation) and electronics noise dominance of smaller loops[27,35,36]. Therefore, developing receiver arrays at 10.5T, the highest field strength currently available for human imaging, requires incorporation of novel decoupling techniques that are effective at 447 MHz (the proton resonance frequency at 10.5T), in order to fully exploit the superior SNR and acceleration potential of the higher magnetic field.

Various radio frequency (RF) array decoupling strategies have been proposed and implemented in the past. Overlap and preamplifier decoupling proposed by Roemer et al.[37] have been improved and used extensively in designs of high density receive arrays[36,38–42]. Noise matching the preamplifiers[27,42] and inductive decoupling[38] were also heavily relied upon in previous works. Furthermore, self-decoupling techniques for low density (up to 8-channels) transmit arrays using intentionally unbalanced capacitive distribution were introduced[43,44]. Lakshmanan et al.[44] proposed the loopole antenna, where segmenting capacitors inside transceiver loops were distributed unevenly to cause an unbalanced current distribution in the loop with electromagnetic field patterns resembling that of a dipole antenna. Yan et al.[43] built on the idea of unbalanced impedances and proposed a transceiver self-decoupling scheme for 7T / 298 MHz where a relatively small RF impedance (e.g. 8.5pF capacitance) is placed opposite a relatively large RF impedance (e.g. 0.4pF) which approaches an open circuit at the RF frequency. This results in the current distribution being unbalanced so that the coil resembles a dipole antenna. However, in contrast to Roemer's work[37], Yan's analysis assumes electric coupling to be limited to coupling via free space and excludes resistive coupling via the conductive sample. While inter-element coupling via the sample can be negligible in the case of loop transmitters far (more than 4 cm away) from the sample, 3D conformal receiver arrays are generally form-fitting and very close to the sample and are designed to be dominated by sample noise[27,28,35]. In this work, we present a strategy for receiver self-decoupling at 447 MHz based on our observation that higher frequencies allow for a more balanced capacitive segmentation of receive elements while maintaining acceptable decoupling. We compare performance of self-decoupled receivers, in terms of SNR, with overlap-decoupled loops. Receiver self-decoupling provides inter-element isolation comparable to overlap decoupling while it does not require geometric overlap or interelement transformers or decoupling networks and is therefore much more practical to implement in high density receive arrays. Furthermore, we show that receiver self-decoupling presented here provides higher SNR compared to overlap decoupling.

Here, we implement the novel receiver self-decoupling technique and build a high density, self-decoupled 32-channel receive array for 10.5T/447 MHz for the first time, without using within-row (axial) overlap[37], explicit inter-element decoupling networks[38,45,46], or highly unbalanced current distributions[43,44], with preliminary results reported in an ISMRM abstract[47]. The self-decoupled 32-channel receive array (10.5T-32Rx) provided substantial experimental peripheral (corresponding to cortical regions in a human head) and central SNR gains compared to an industry-standard 32-channel receive array at 7T. Furthermore, parallel imaging performance at 10.5T was superior to that of 7T, with 10.5T-32Rx providing acceleration performance comparable to a 64-channel array coil at 7T.

## 2. METHODS
### 2.1. 16-Channel Transmitter

The primary focus of this paper is on the contribution of receiver technology and $B_0$ field strength to SNR and parallel imaging performance. Therefore, the transmitter design is covered here only briefly, with its detailed characterization published separately[48]. A 16-channel transmitter comprising two rows of 8 inductively decoupled rectangular loops was used[39,48]. The 2-row design of the transmitter array[39,49,50] has the potential of increasing degrees of freedom in parallel transmit (pTx) RF pulse design especially for specific absorption rate (SAR) control[51–53]. In order to minimize transmit-receive interaction, the transmitter was actively tuned, i.e. a PIN diode circuitry was used to tune the transmitter during signal transmission, leaving it off-resonance during reception. 10.5T is currently operated under an FDA Investigational Device exemption (IDE) which requires that all RF coils planned for human use are approved by the FDA. The procedure to obtain necessary approvals for in-vivo human brain imaging using this transmitter together with a receive coil inserted in it is currently ongoing.

**2.2. Receiver Self-Decoupling**

Rectangular loops similar in size to those used in the array were modeled in several scenarios. First, two 5x5cm² loops and then, two 2.5x5cm² loops were positioned (in simulations) on a flat surface in proximity to a cubic phantom (permittivity $\epsilon_r = 50$, conductivity σ = 0.6S/m to approximate human brain tissue properties[54] across all simulations). These loops were constructed using 18 AWG copper wire (diameter 1.02mm) and were divided into four segments using three capacitors and the feed point. Two fixed capacitors and a trimmer capacitor ($X_{tr}$), with values in the same range as fixed capacitors, were used inside each loop. The feed circuitry, presented in Figure 1c, consisted of a detune trap as well as tune ($X_t$) and match ($X_m$) adjustable capacitors. Similar principles were used to construct non-overlapped 10x10cm² loops using balanced capacitive distribution on a cylindrical surface at a constant distance from a cylindrical sample ($\epsilon_r = 50$, σ = 0.6S/m) as shown in Figure 2b.

Electromagnetic/circuit co-simulations were performed using CST Studio (Dassault Systèmes Simulia Corp., Johnston, RI). Coil elements were modeled in SolidWorks (SolidWorks Corp., Waltham, MA) and imported into CST. Simulations were performed over a frequency range of 2 GHz using the finite-difference time-domain method to solve Maxwell's equations and were partially accelerated using one GPU. The loops were tuned to 447 MHz (proton resonance at 10.5T) and matched to 50Ohm. The simulation pipeline has been discussed in more detail in previous works[48].

Several electromagnetic (EM)/circuit optimization problems were set up with the goal of finding $X_{tr}, X_t, X_m$ to minimize $S_{21,f}$ conditioned on $S_{11,f} \leq -12$dB and $S_{22,f} \leq -12$dB where f = 447 MHz as the resonant frequency. Values of $X_{tr}, X_t, X_m$ were limited to practical ranges guided by bench experiments. Other 3D EM model parameters, including the distance between loop elements and the gap between the coils and the phantom, were kept constant and equal to practical receive array values during optimization. Note that the objective function and the cost (loss) function for these optimizations can be defined in various ways. In our experience, incorporating the objective ($S_{21,f}$) into the loss function to form a weighted-linear combination of individual L1 norms in the linear (as opposed to dB) scale of real and imaginary parts of the S-parameters resulted in faster convergence. The trust region method and the Nelder Mead simplex algorithm were used to solve for optimums. It should be noted that in practice (on the bench) such an optimization is straight-forward and is substantially less time consuming compared to the numerical simulations because, on the bench, the results of parameter modifications can be monitored immediately in an analogue manner using a vector network analyzer.

Numerical results for S-parameters and per-port, complex-valued H-fields at 447 MHz were exported to ASCII files. Post-processing and analysis of magnetic field results were performed using customized Python scripts. Complex-valued receive magnetic fields were calculated for each coil element using $B_1^- = \mu_0(H_x^* + jH_y^*)/2$ where μ₀ is the permeability of free space and $x, y$ are Cartesian coordinates orthogonal to the static magnetic field ($B_0$). In order to compare the receive field between overlap and self-decoupling methods, two metrics were calculated: sum of magnitude (SOM) of the complex combined magnetic fields $\text{SOM} = \sum_{\text{sample}} |\sum_{channels} B_1^-|$ and the root sum of squares (RSOS) of magnitudes of each channel's magnetic field $RSOS = \sum_{sample} \sqrt{(\sum_{channels} |B_1^-|^2)}$, in other words $RSOS = \sum_{sample} \sqrt{B^H B}$ where $B$ is a vector composed of complex magnetic receive fields of each individual channel and $H$ is Hermitian transpose. Furthermore, RSOS was corrected for noise correlation to calculate noise-correlation-weighted SNR as given by $SNR = \sum_{sample} \sqrt{B^H \psi^{-1} B}$ where $\psi$ is the normalized noise correlation matrix calculated using the simulated complex-valued scatter matrix[19,27,55–57]. The summation over sample is intended to collapse spatial maps into a single numeric metric for comparison purposes. These metrics are particularly appropriate here as we are considering receive-only coils, so we are interested in SNR, not transmit efficiency.

### 2.3. 32-Channel Receiver

A close-fitting receive former (helmet) was designed while considering physical constraints imposed by the transmitter, the dimensions of which were directed by a 41 cm ID head gradient coil that will be used in the future. The shape of the former was optimized based on previous helmet designs, numerical model of a human head, and consideration of average range of head sizes. A structure for mechanical support of the preamplifiers and cables was designed to be mounted as an additional part on top of the head former (Figure 1b). A visual channel at the level of the eyes was designed to accommodate the needs of future fMRI studies, which often involve visual stimulus presentation or eye tracking. The distribution of preamplifier substrates was driven by cable routing design, which will be discussed subsequently, as well as preamplifier interactions and directional sensitivity relative to the $B_0$ direction.

The 32Rx receive array is composed of 31 loops, divided into four rows along the z-axis, and one cloverleaf element covering the top. The layout of the loops, presented in Figure 1a, allowed for partial overlap among elements of the different rows (8-10 mm) and gaps between neighboring loops within each row (5-8 mm); the self-decoupled loops with gaps in the axial (x-y) plane are included here to provide an SNR advantage compared to overlapped loops (see Table 1). This results in a high-density receive array designed to rely primarily on self-decoupling. Each row of loop elements was shifted (by a half loop) compared to the neighboring rows (i.e. rotated azimuthally in the real coil). This self-decoupled design substantially reduced the construction complexity that would arise from overlapping adjacent loops within each row or using decoupling networks or transformers between loop elements and contributed to the SNR improvements. Overlap along the z-axis (along the center of the MR scanner bore) was maintained to further improve SNR via increased channel density. A key consideration in designing the layout was a primary focus on the visual cortex, driven by a large array of vision neuroscience projects conducted at CMRR that can greatly benefit from higher field strength; this, in turn, motivated an increase in density in the posterior array for the rows that will be facing the occipital lobe, at the expense of a reduced density at the top (six loops) and bottom (three loops) rows. Loop elements with the plane of the loop aligned perpendicularly to the z-axis would have compromised sensitivity; therefore, a cloverleaf element, rather than loop elements, was placed at the top of the coil, resulting in a Poynting vector perpendicular to the z-axis. The cloverleaf element was composed of two figure 8 loops placed perpendicular to each other; the outputs of the figure 8 elements were combined in hardware as a quadrature pair to form a single receive channel.

A prototype composed of eight channels, arranged in four rows to be representative of the final layout, was initially built and tested on the bench prior to measurements in a head shaped gel phantom at 10.5T. A 16 AWG silver-coated wire was used to construct the loops. Lumped capacitor values employed in the loops were 3.3pF, 4.7pF, and 6.8pF. One trimmer capacitor with a value range of 8-20pF or 2-6pF (SGC3S300NM or SGC3S060NM, Sprague-Goodman, NY, USA), included inside each loop, was carefully adjusted to decouple the loops in each row based on their scatter matrix parameters. The values of the larger trimmers were measured to be in the range of 8.5-15pF after adjustment. The feed board, including the active detuning circuitry, and the preamplifier board were similar to those presented in a previous ISMRM abstract[58].

### 2.4. Receiver Noise Correlation Mitigation

On the bench, the scatter matrix ($S_{ij}$) was measured between all coil pairs using the 8-channel prototype. Similar $S_{ij}$ measurements for the completed 32-channel array were limited to adjacent coils in the same row or overlapping coils from different rows, considered to represent worst-case scenarios. Measurements were done after tuning, matching, and self-decoupling the coils and both prior to and after adding preamplifiers. Self-decoupling of adjacent loops within each row provided robust inter-element isolation. Low noise preamplifiers (WMA447A, WanTcom Inc., Minneapolis, MN) with input impedance of 1.5Ω and noise figure of 0.45dB were used for preamplifier decoupling. Preamplifiers were mounted on a ground plane which helped minimize interactions with transmitter/receiver elements and preamp oscillations. To reflect the actual use case, $S_{ij}$ measurements were done with all receive elements in the tuned state.

Preamplifier interactions and coaxial cable interactions were investigated to characterize the effects of stacking preamps/coaxial cables close to each other as well as cable routing. Noise correlation matrices for various cabling, preamplifier configurations, and cable trap locations were measured inside the 10.5T MR scanner with the 8-channel prototype and a head shaped phantom. Coaxial cables were isolated using self-shielded input cable traps to suppress shield-current-induced noise[59,60]. Traps were carefully tuned after being installed on preamplifier input coaxial cables; however, several cable traps were intentionally tuned slightly off-resonance to mitigate their interaction with resonant loops of the receiver or transmitter[39]. These input cable traps are resonant structures constructed using a trimmer capacitor (8-30pF, SGC3S300NM, Sprague-Goodman, NY, USA) soldered to copper tape wrapped around a dielectric cylinder mounted on the outer conductor of preamplifier input coaxials (with the copper tape soldered to the outer conductor). One further s-matrix measurement was made with output cable traps detuned to investigate potential crosstalk between those resonant structures as well.

### 2.5. Transmit-Receive Interaction

Design of routing paths for preamplifier input coaxial cables was driven by noise correlation and receive-transmit interaction considerations. Coaxial cables may present a significant conductor barrier to the transmitter. Coaxial cables longer than 1/10[th] wavelength (6.7cm at 10.5T) have considerable electromagnetic interaction, significantly distorting transmit magnetic field ($B_1^+$) and adversely affecting the receive array noise correlation matrix[61]. At 7T, or 298 MHz, 1/10[th] of the wavelength is 10cm; as such, input coaxial cables (8-9cm) would not be as electromagnetically problematic as they are at 10.5T. Based on previous experience with loop transmitter designs[38], knowledge of transmit field patterns[48], and prototype experiments explained above, it was determined that collecting coaxial cables in five paths along the z-axis parallel to the center of the transmit loops would result in minimum Tx/Rx interaction.

Measurements of the receive-transmit interaction were performed both on the bench and in the scanner. On the bench, scatter matrix parameters of the transmitter were monitored before, during, and after insertion of the receive array, while the receiver was actively detuned using DC supply to PIN diodes[58] and loaded with a human head shaped phantom. In the scanner, transmit field maps were acquired in two configurations: first, with the 16-channel transmitter and 32-channel receiver as an ensemble of transmit-only and receive-only arrays, second, with the 16-channel transmitter used as a transceiver in the absence of the 32-channel receiver. Relative transmit $B_1^+$ maps were acquired using a small flip angle multi-slice gradient echo (GRE) sequence with magnitude images from sequential single channel transmissions normalized by their sum[62]. Absolute transmit $B_1^+$ maps were then generated by normalizing relative transmit maps by $sin(\alpha(\vec{r}))$ where $\alpha(\vec{r})$ is the actual flip angle as a function of spatial coordinate $(\vec{r})$ obtained via 3D GRE Actual Flip Angle (AFI) acquisitions obtained with all channels transmitting[62–64]. Absolute transmit field maps were acquired both with and without the receive array in place, with the transmitter used as a 16-channel transceiver when the 32-channel receive array was not inserted.

### 2.6. SNR and g-Factor

*Data acquisition*

Data were acquired on a 10.5T MRI (Siemens Healthcare, Erlangen, Germany) system using a phantom with permittivity $\epsilon_r \simeq 47.25$, conductivity $\sigma \simeq 0.65 S/m$ at 447 MHz as measured with a DAKS-12 (SPEAG, Zürich, Switzerland) to approximate human brain tissue properties[54]. Phantom dimensions are provided in Figure 7a. All SNR measurements were replicated two times, once with the 16Tx/32Rx coil described above at 10.5T, then with an industry-standard 1Tx/32Rx coil (Nova Medical Inc., Wilmington, MA) at 7T using protocols, acquisition parameters, setups, and data analysis pipelines similar to those at 10.5T.

Relative transmit field maps were obtained using a series of small flip angle multi-slice gradient echo (2D GRE) sequences, pulsing on one transmit channel at a time[65,66]. Typical sequence parameters were seven 5 mm thick axial slices (with center slice being isocenter), TR=100ms, TE=3.5ms, Flip Angle (FA)=10°, pixel bandwidth = 300 Hz/pixel. Actual flip angle (AFI)[64] maps were acquired (3D, TR=75ms, TE=2.0ms, nominal FA=50°) with all channel transmitting together. SNR data were acquired at approximately fully relaxed longitudinal magnetization, with a multi-slice GRE sequence, with TR=7000ms-10000ms, TE=3.5ms, FA=80°, pixel bandwidth = 300 Hz/pixel. This was followed by a *noise* scan which was identical to the *SNR* sequence except for FA=0° (no RF pulse), TR=70-100ms[67]. When measuring AFI maps for SNR normalization at 10.5T, a 16Tx, circularly polarized (CP-like) transmit field $B_1^+$ phase shim setting was calculated for the 16 transmit channels allowing for acceptable $B_1^+$ efficiency over a large fraction of the phantom[26,68].

*SNR and g-factor calculations*

The noise correlation matrix was calculated based on complex noise data (obtained in the absence of RF pulsing) and used to decorrelate the SNR data before they were combined using root sum-of-square method[37,56]. In the steady state, signal intensity in a GRE sequence is proportional to $\rho(1 - E_1)(sin\,\theta)/(1 - E_1 cos\,\theta)E_2$ where ρ represents proton density, $E_1 = exp(-T_R/T_1)$, $E_2 = exp(-T_E/T_2^*)$ and θ is the spatially varying, voltage-normalized actual flip angle. With $T_R \gg T_1$ it follows that $E_1 \ll 1$ which results in $SNR \propto \rho\,sin(\theta)\,E_2$. SNR maps were normalized by $sin(\theta)$ derived from AFI maps, voxel size, number of acquisitions, number of samples along the read-out and phase-encoding directions, and bandwidth to make SNR calculations comparable across experiments[63]. SNR ratios were further normalized by $T_2^*$ values which were measured at 34ms and 28ms at 7T and 10.5T, respectively. The noise figure from receiver chain of the MR scanners (excluding RF coil and preamplifiers) was not reflected in these calculations. Noise amplification in accelerated images was quantified in terms of g-factor[30] and calculated as $g =$ $SNR/(SNR_R \times \sqrt{R})$ where R is the acceleration factor and $SNR_R$ is the accelerated SNR calculated based on fully sampled acquisitions that were retrospectively under-sampled. In order to be able to compare g-factor numbers of the 10.5-32Rx presented here with previously published[23] g-factors of 7T-32Rx and 7T-64Rx, the same acquisition and post-processing pipeline used for 7T-32Rx and 7T-64Rx were maintained in our experiments with the 10.5T-32Rx.

## 3. RESULTS
### 3.1. Receiver Self-Decoupling

Figure 2(b-e) presents simulation results for two 10x10cm² receive loops. The resulting crosstalk between receive elements is $S_{21} = -12$ dB. Figure 2d shows the magnetic field $B_1$ of a single excited loop which is consistent with the classic field pattern of a loop at this frequency, implying sufficient isolation. The surface current distribution is almost symmetric with strong current distributed on all segments of the excited loop (and limited current induced on the second loop) (Figure 2e), again demonstrating effective interelement isolation

Simulation results for 25x50mm² receive loops demonstrated -11.2dB isolation using the self-decoupling method (compared to -13.8 dB using overlap) at 447 MHz. In the case of 50x50mm² loops, self-decoupling provided $S_{21} = -12.9$dB interelement isolation (compared to -16.8dB using overlap). These results indicate that the proposed self-decoupling method achieves better than -11dB isolation (in the worst-case scenario), which is sufficient for receive array design given that preamplifier decoupling will further improve the isolation.

Figure 3 depicts the $B_1^-$ resulting from two self-decoupled loops and provides a qualitative comparison with $B_1^-$ from two loops of the same size but overlapped. The RSOS combination of receive signals from the two self-decoupled loops shows strong receive signal despite the gap between the two loops (Figure 3a); a hint of the separation between the two loops is present in the form of two overlapping distributions of $B_1^-$ with two distinct maximal penetration peaks. The two overlapping loops, as expected show more of an overlapping $B_1^-$ distribution. Table 1 presents a quantitative comparison between SOM and RSOS of $B_1^-$ and SNR integrated over the sample of self-decoupled and overlapped loops for two sets of loop sizes. Compared to overlapped loops, the proposed self-decoupling method provides 17% more RSOS $B_1^-$ for the 25x50 mm² loops and 32% more RSOS $B_1^-$ for the 50x50 mm² loops. Furthermore, self-decoupling results in 10% and 26% higher noise-correlation-weighted SNR for 25x50mm² and 50x50mm² loops, respectively.

### 3.2. Receiver Noise Correlation and Interaction with Transmit Field

The inter-element isolation was measured prior to preamplifier decoupling as $S_{21}$ at the resonance frequency of 447 MHz, using a calibrated 16-port vector network analyzer (VNA, ZNBT8, Rohde & Schwarz); these values were on average in the range of 11-12dB for self-decoupled adjacent coils in the same row (see Figure 2a), 12-15dB for partially-overlapped coils from different rows and in the range of 20-30dB for distant neighbors, demonstrating effective self-decoupling of the 32-channel receive array without the need for overlap in each row, explicit transformer decoupling, or unbalanced capacitive distribution. Preamplifier decoupling further reduced crosstalk between receive array elements to negative 35-40dB.

The noise correlation matrix measured inside the 10.5T MR scanner resulted in maximum noise correlation of 0.37 (Figure 4), which is a significant improvement compared to previous works[23]. We attribute this to the novel self-decoupling technique, experimentally optimized cable routing (relative to transmit elements to minimize shield-current-induced noise and transmit field distortion) and cable trap locations (to minimize trap interference with receive element resonances).

Figure 5 shows power-normalized transmit field maps (in $\mu T/\sqrt{W}$) measured in the 10.5T scanner using the 16-channel transmitter with and without insertion of the 32Rx receive array, with the phases of the different channels on the transmitter set so as to generate a circularly polarized $B_1^+$ field, which results in the center bright $B_1^+$ maps[26]. These maps demonstrated less than 10% distortion of the transmit field upon inserting receive array; as such, characterizing the limited transmit field change following the insertion of the receive array. This characterization allows for streamlining safety validation studies by obviating the need for inclusion of the receive array in electromagnetic simulations of specific absorption rate (SAR).

### 3.3. SNR and g-Factor

Figure 6 illustrates experimental unaccelerated SNR comparisons in axial slices obtained with identical protocol, experimental setups, and analysis pipeline using a 32-channel receiver at 7T (7T-32Rx) and a 32-channel receiver at 10.5T (10.5T-32Rx). The phantom and region of interest (ROIs) used for SNR comparison are depicted in Figure 7(a-b). SNR comparisons were made by averaging over a central region and a peripheral region defined by a circular boundary at depth "d" into the phantom (Figure 7b). Table 2 presents the SNR ratios for various peripheral *versus* central segmentations parametrized by the depth "d" of the central/peripheral boundary. SNR comparison demonstrated 1.46 times central and 2.08 times peripheral SNR gains at 10.5T with *d = 20* mm. Figure 7c provides unaccelerated 3D (axial, coronal, sagittal) SNR comparisons between 10.5T-32Rx and 7T-32Rx in the central slices.

The overall average SNR gain using the 10.5T-32Rx was 81%. Expressed as an exponent *n* of the static magnetic field strength (i.e. as $B_0^n$), the overall average $n \simeq 1.5$, and in the periphery $n \simeq 1.8$ (using SNR ratios for *d = 20* mm). This enhanced SNR is primarily attributed to higher static magnetic field $B_0$ strength as well as noise mitigation and SNR advantages provided by the self-decoupling method (see Table 1). The less-than-expected SNR gains at the center are probably due to smaller loop sizes and non-overlapping loops employed in self-decoupling, resulting in lower density of azimuthally distributed coverage. For a given row of azimuthally distributed loops, overlapping the loops would lead to either a larger number of loops (when loop size is equal to the self-decoupled design), or radially larger loops (when loop count is equal to the self-decoupled design). Both cases may produce a stronger receive B1 field in the central regions.

The performance of the 10.5T-32Rx in 2D accelerated acquisitions can be compared with the 7T-32Rx in terms of g-factors (presented in Table 3). The mean inverse g-factor of the 10.5T-32Rx (1/g = 0.69) is 18% more than mean inverse g-factor of the 7T-32Rx (1/g = 0.59) for 4x4 acceleration.

### 4. DISCUSSION

Previous studies have shown that ultimate intrinsic SNR increases with the magnetic field magnitude[16,18,31]; the use of high density receive arrays is necessary at UHF to approach ultimate intrinsic SNR[28] available at ultra-high fields and capitalize on the acceleration potential of ultra-high field MRI. However, capturing these potential gains experimentally with increasing number of channels in UHF also depends in the execution of the RF coil array, both in accelerated and unaccelerated imaging. As such, strategies for mitigating interelement coupling and receiver noise play a critical role.

Here, we improved upon a novel self-decoupling approach[43] to build a 10.5T 32-channel receive array. The self-decoupling method presented by Yan et al.[43] introduced a promising new method for decoupling of high density receiver arrays. However, the method at 3T and 7T operating frequencies requires significantly unbalanced distributed capacitors or introduction of inductive circuit elements to achieve the desirable unbalanced current distributions. The same self-decoupling effect was achieved at 10.5T with a more uniform capacitive distribution for the loop sizes appropriate for a 32 channel receive array targeting coverage over the human brain.

The effort presented here is the first implementation of the self-decoupling method in a 3D conformal, high density 32-channel receive-only array for human brain imaging, albeit in this case it was done for the very high frequency of 447 MHz. Our data confirms Yan et al.'s suggestion that in high-density, 3D conformal receive arrays, self-decoupling can be paired with preamplifier decoupling to improve inter-element isolation. Importantly we demonstrate that the combination of higher frequency (447MHz), smaller loop sizes appropriate for a 32-channel receiver and distributed inductance of such loops being in the range required for self-decoupling, can be exploited to achieve acceptable inter-element isolation with a more uniform capacitive distribution at 447MHz. The resulting approach improves the RSOS of magnitudes of receive fields substantially, and results in better noise-correlation-weighted SNR compared to overlapped decoupling.

Experimental noise correlation matrix demonstrated effective inter-element receiver decoupling which is indicative of promising potential for parallel imaging. Indeed, the array provided substantially improved parallel imaging performance at 10.5T compared to a 32-channel industry standard RF coil operating at 7T, and comparable to a 64-channel array at 7T (Table 3). It is not possible to distinguish if the gains are primarily because of the coil design and inter-element decoupling approach employed or reflect the gains expected from going to a very high magnetic field[19,22,31]. Nevertheless, the approach employed has enabled the realization of the anticipated gains in parallel imaging performance at UHF that push the electromagnetic operating regime toward the far field domain.

We anticipate that at ultra-high frequencies (447 MHz and higher) the self-decoupling strategy will significantly simplify future high density receive array design and construction as it obviates practical complexities of common decoupling techniques. In addition, the receive array as constructed was shown to result in limited transmit field distortion following insertion into a complex, 16-channel transmitter. This will facilitate the electromagnetic simulation effort for specific absorption rate (SAR) and regulatory validation by accounting for the small impact of the receive array by including it in a safety margin when calculating the maximal allowable power use. Furthermore, limited receiver-transmitter interaction should enable an interchangeable coil setup where a single transmitter can be used with 32-channel, 64-channel, and 128-channel receivers.

Peripheral and central SNR gains presented here in comparison to 7T confirm the expected gains for ultra-high field imaging targeting the human head[18,69–72]; however, consistent quantitative comparisons are difficult to make. In EM simulations, it is generally reported that the gains in ultimate intrinsic SNR are supralinear with field magnitude in the center of the head[18,19,70], all suggesting approximately quadratic increases with field magnitude. In the periphery, however findings are not so consistent; Wiesinger et. al.[19] still predict supralinear gains in the periphery but other studies predict linear[18] or even less[70]. In part, the discrepancies may emerge from the particular human head model[18,70], or approximations of the human head[70] employed, physical locations of the current distribution relative to the model, and the ROI's over which the ultimate intrinsic SNR was averaged. Irrespective of the quantitative differences, however, the afore mentioned studies predict supralinear gains with field magnitude averaged over the brain, approximately in agreement with $B_0^{1.65}$ dependence experimentally reported by Pohmann et. al.[69]. The results from our 7T and 10.5T comparison agrees with this general conclusion. However, the gains we measured were less than quadratic overall, while they were quadratic in the peripheral ROI. In interpreting these results, it must be realized that different receive array designs can lead to different SNR gain distributions and may have different degree of success in capturing the ultimate intrinsic SNR in different regions of the object being imaged; this will impact comparisons between two different coils at two different field strengths. Moreover, that the implementation of a real coil can deviate significantly from the modeling efforts, especially of ultimate intrinsic SNR. Compared to simulations of ultimate intrinsic SNR, the proximity of the coil elements to the object is not uniform, nor is the size of the coils employed. While we demonstrate that the self-decoupling approach is an improvement in capturing SNR compared to the overlapped-decoupling method when both methods employ equal numbers of geometrically identical loops, an overlapped design allowing for larger loops covering the same surface area may potentially improve central SNR. There are also other potential noise sources, such as imperfect preamplifier decoupling which can lead to the preamplifier noise being propagated back into receive loops.

This work is a significant milestone also towards building 64-channel and 128-channel receiver arrays for human brain imaging at 10.5T. It is fully anticipated that scaling up to a 128-channel design will pose additional challenges in noise mitigation, preamplifier design, and receiver-transmitter interactions. However, expected SNR and parallel imaging gains provide strong rationale and impetus to address such challenges.

## 5. CONCLUSIONS

There is significant clinical and research interest in capitalizing on the acceleration and signal-to-noise ratio (SNR) potential of ultra-high field MRI. Here, we present the first self-decoupled 32-channel receive array (32Rx) and capture substantially superior SNR and parallel imaging performance using a 10.5T MRI system compared to a 7T MRI system. The 10.5T-32Rx provided significant peripheral and central SNR gains compared to an industry-standard 7T-32Rx, both in unaccelerated and 2D accelerated acquisitions. This achievement delivers the much-anticipated SNR boost in highly accelerated ultra-high field imaging required for further understanding of human brain function and connectivity.

**ACKNOWLEDGMENT**

The authors acknowledge the constructive feedback of Myung Kyun Woo, Lance DelaBarre, Yigitcan Eryaman, and Matt Waks (all with Center for Magnetic Resonance Research, University of Minnesota, Minneapolis, MN 55455) and funding from NIH U01EB025144, P41 EB027061, NIH S10 RR029672, and NIH R34AG055178 grants.

**FIGURES & TABLES**

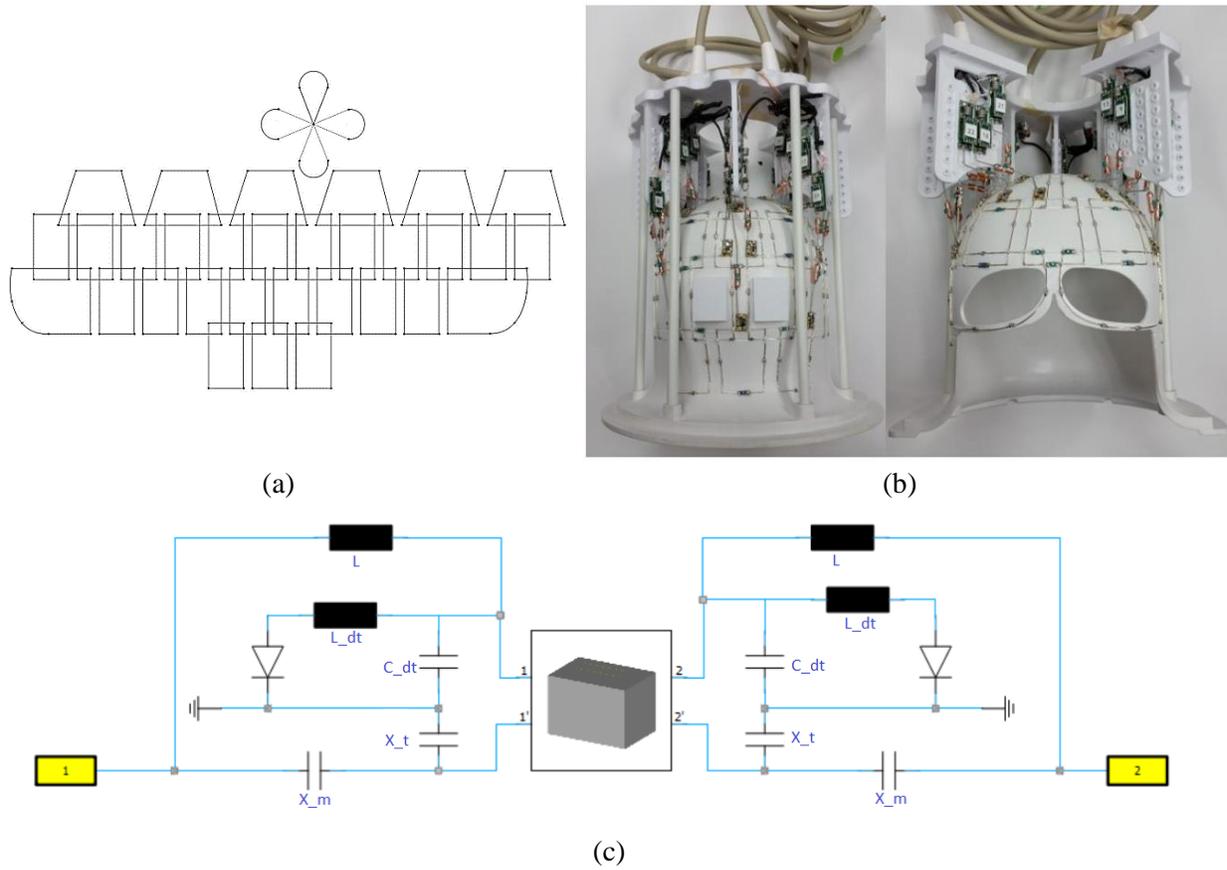

Figure 1 – The 10.5T 32-channel receive array, (a) flattened layout of the channels, (b) as-built image of the coil, (c) feed circuit (for two channels) consisting of tune ($X_t$) and match ($X_m$) capacitors as well as detune trap ($L_{dt}$ and $C_{dt}$). The trap is used to detune the receiver during transmit.

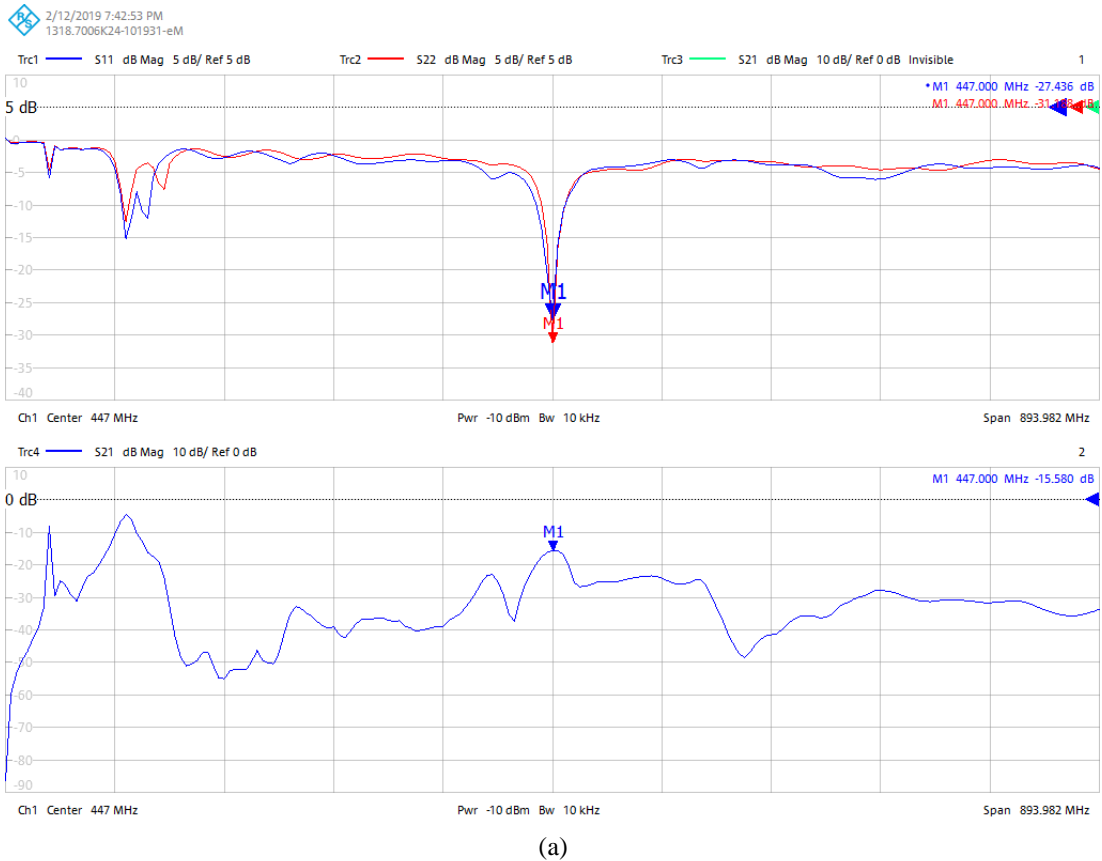

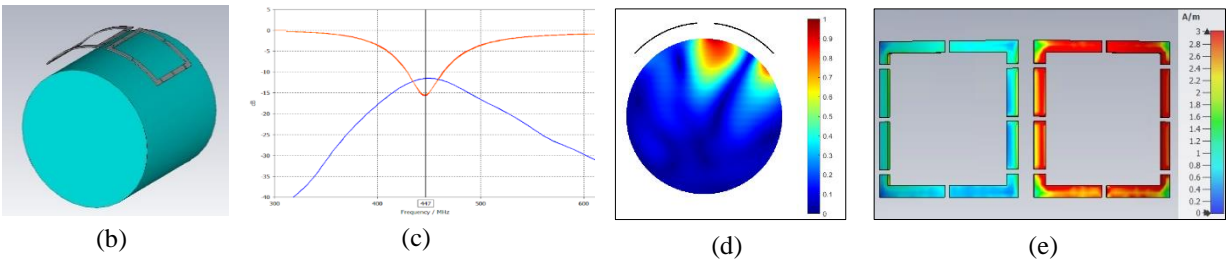

Figure 2 – (a) Experimental VNA screenshot demonstrating effective self-decoupling of adjacent loops with S11= -27dB, S22= -31dB, S21= −15dB (including 2dB cable losses) at 447 MHz; (b – e) Simulation results for two self-decoupled rectangular loops at 10.5T/447 MHz. (b) Electromagnetic model of the loops loaded at 10.5T/447 MHz, (c) S11=S22= -15dB (red) and S21= -12dB (blue) at 447 MHz, (d) magnitude B1 field of one loop (axial, center slice, normalized to 1) showing the location of the loops as arcs, (e) average absolute surface current distribution (when only the right loop is excited).

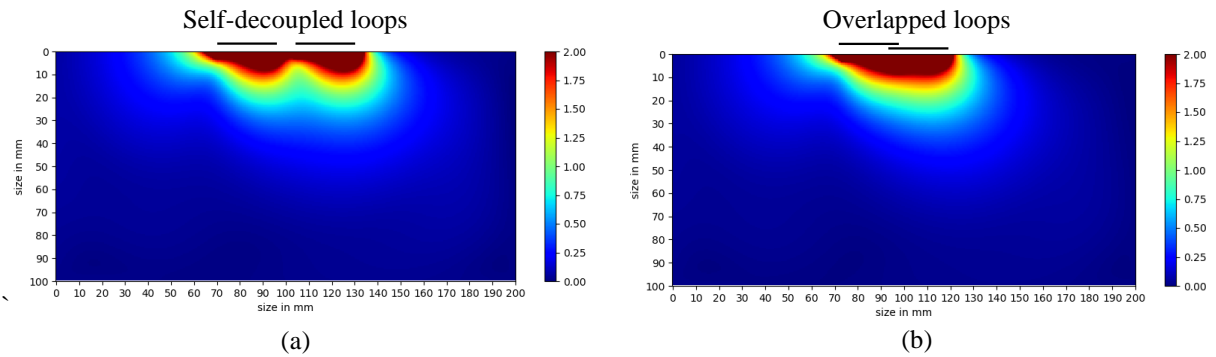

Figure 3. Normalized combined $B_1^-$ receive fields for two rectangular 25x50mm$^2$ loops, with (a) root-sum-of-squares (RSOS) of receive fields for self-decoupled loops, and (b) RSOS for overlapped loops. The black lines delineate the location of the loops relative to the phantom.

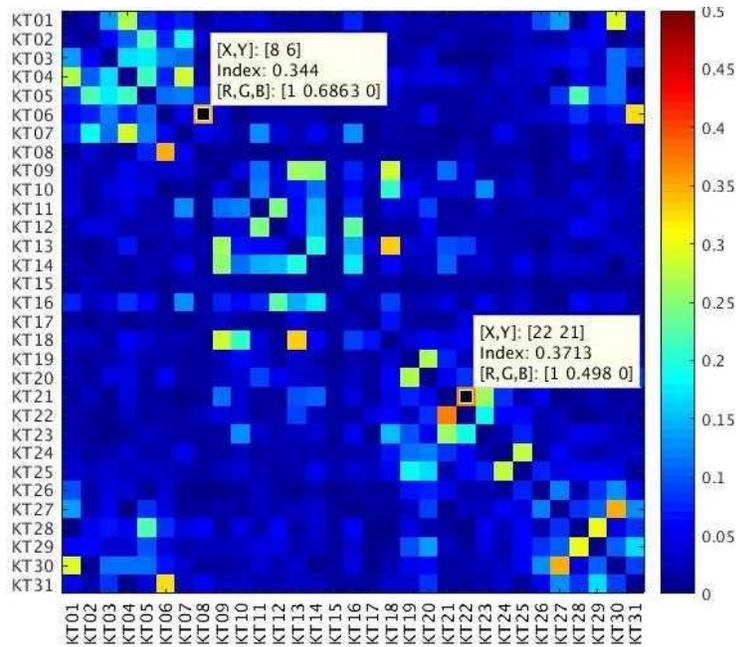

Figure 4 – Experimental noise correlation map for the 10.5-32Rx receiver demonstrating effective decoupling of receive elements. Maximum noise correlation is 0.37.

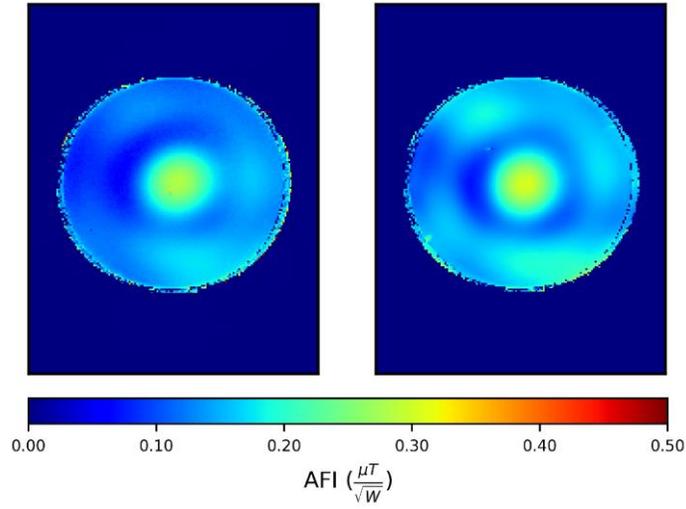

Figure 5 - Comparing experimental normalized transmit field maps of (left) 10.5T-16Tx used as transceiver without the presence of 32Rx versus (right) 10.5T-16Tx/32Rx demonstrating the effect of Rx insertion to be limited to $\pm 10\%$ where signal intensity is reliable.

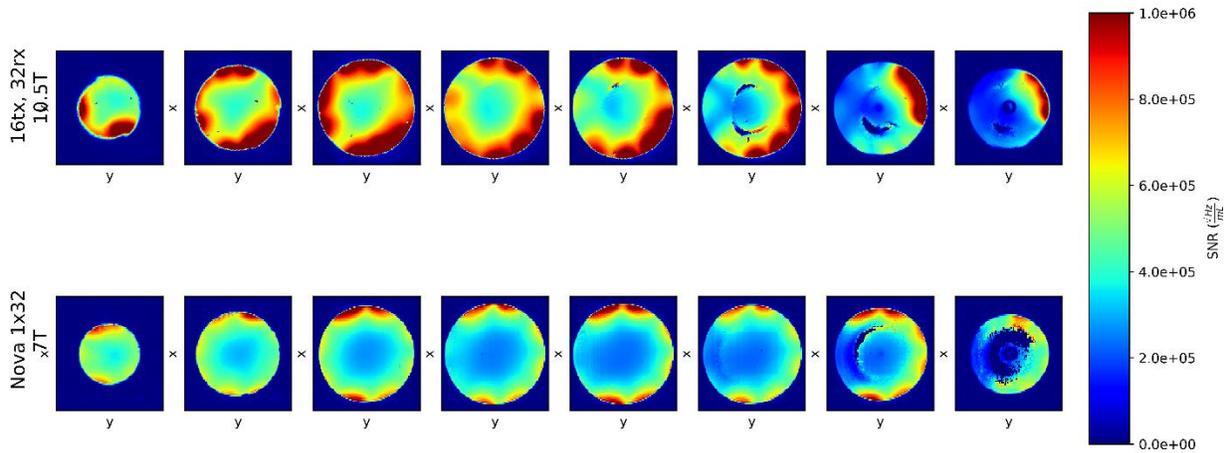

Figure 6 – Comparing normalized SNR of the 10.5T-32Rx (top row) and industry-standard 7T-32Rx (bottom row) on eight axial slices. Local defects (the crescent-shaped signal loss in first three slices from right) are due to specific transmit $B_1^+$ shim. The dark center (in the first two slices from right) is an oil reference. Numbers in $\sqrt{(Hz)}/ml$.

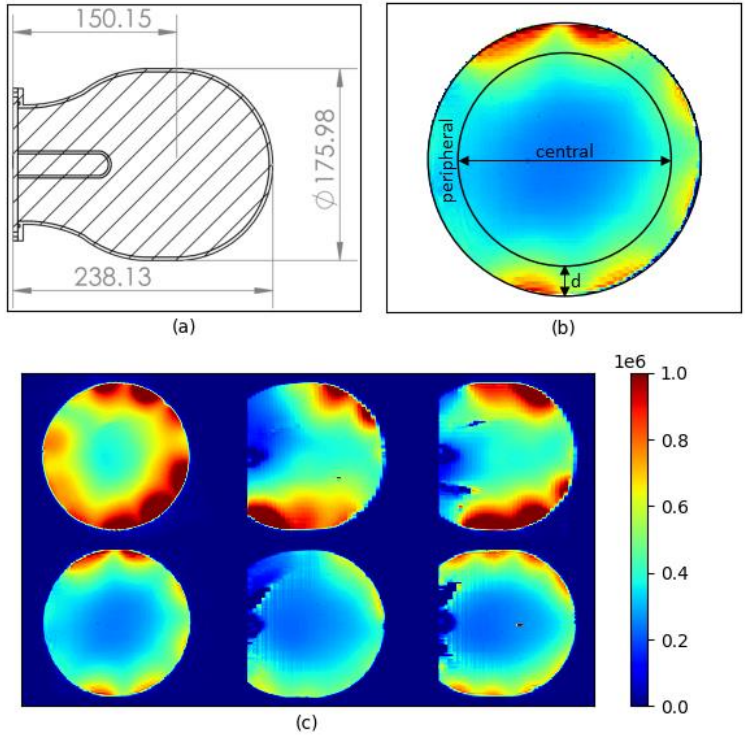

Figure 7 – (a) phantom cross section; the cylindrical tube is an oil reference. (b) axial cross section of the phantom, showing the central versus peripheral regions demarcated by $d$, depth into the sample. (c) comparing SNR from the 10.5T-32Rx (top row) with SNR from 7T-32Rx (bottom row) in three orthogonal planes (SNR in $\sqrt{(Hz)}/ml$).

| Loop size | SOM | RSOS | SNR |
|---|---|---|---|
| 25x50 mm$^2$ | 19 | 17 | 10 |
| 50x50 mm$^2$ | 47 | 32 | 26 |

Table 1. Percentage gain in SOM, RSOS, and SNR obtained using self-decoupled loops instead of overlap-decoupled loops (calculated as (self_decoupling_snr – overlap_snr)/overlap_snr * 100) at 10.5T/447 MHz.

| Depth (d, mm) | Peripheral SNR ratio | Central SNR ratio | Overall Average SNR ratio |
|---|---|---|---|
| 30 | 1.95 | 1.44 | 1.81 |
| 25 | 2.00 | 1.46 | 1.81 |
| 20 | 2.08 | 1.46 | 1.81 |
| 15 | 2.25 | 1.47 | 1.81 |
| 10 | 2.58 | 1.47 | 1.81 |

Table 2 – Ratio of the average 10.5T-32Rx SNR divided by average 7T-32Rx SNR for various values of d (peripheral vs central boundary, delineated in Figure 7b).

|  | g-factor |  | Inverse g-factor |  |
|---|---|---|---|---|
| 10.5T-32Rx | Mean | 1.44 | Mean | 0.69 |
|  | Max | 1.8 | Min | 0.56 |
| 7T-64Rx | Mean | 1.3 | Mean | 0.77 |
|  | Max | 1.9 | Min | 0.53 |
| 7T-32Rx | Mean | 1.7 | Mean | 0.59 |
|  | Max | 2.7 | Min | 0.37 |

Table 3 – Comparing g-factor values for 4x4 2D accelerated acquisitions using 10.5T-32Rx with g-factor values for 7T-64Rx[23] and 7T-32Rx[23].